\documentclass[9pt,twocolumn,twoside]{osajnl}

\journal{pr} 
\usepackage{graphicx}
\usepackage{dcolumn}
\usepackage{lipsum}
\usepackage{bm}
\usepackage{tabu}
\usepackage[utf8]{inputenc}
\usepackage[T1]{fontenc}
\usepackage{xcolor}
\usepackage{ amssymb }

\usepackage{comment}
\usepackage{soul}

\title{Free--space non--separability decay of clasicaly--entangled modes}

\author[1]{Xiao-Bo Hu}
\author[2,*]{Benjamin Perez-Garcia}
\author[3]{Valeria Rodr\'iguez-Fajardo}
\author[2]{Raul I. Hernandez-Aranda}
\author[3]{Andrew Forbes}
\author[1,4,$\dagger$]{Carmelo Rosales-Guzm\'an}
\affil[1]{Wang Da-Heng Collaborative Innovation Center for Quantum manipulation \& Control, Harbin University of Science and Technology, Harbin 150080, China}
\affil[2]{Photonics and Mathematical Optics Group, Tecnologico de Monterrey, Monterrey 64849, Mexico}
\affil[3]{School of Physics, University of the Witwatersrand, Private Bag 3, Johannesburg 2050, South Africa}
\affil[4]{Centro de Investigaciones en Óptica, A.C., Loma del Bosque 115, Colonia Lomas del campestre, 37150 León, Gto., Mexico}
\affil[*]{b.pegar@tec.mx}
\affil[$\dagger$]{carmelorosalesg@harbust.edu.cn}

\begin{abstract}
One of the most prominent features of quantum entanglement is its invariability under local unitary transformations, which implies the degree of entanglement remains constant during free-space propagation. While this is true for quantum and classically--entangled modes, here we demonstrate a novel type of classically-entangled modes that experience an entanglement decay upon free-space propagation. We show this by numerical simulations and corroborate experimentally. Our results evinces novel properties of classically-entangled modes, which pave the way to novel applications.
\end{abstract}

\setboolean{displaycopyright}{true}

\begin{document}

\maketitle
\section{Introduction}
It is well known that entanglement is invariant to local unitary transformations. An example of such is the well-known fact that the degree of entanglement remains unmodified upon free-space propagation. Crucially, non-separability, the fundamental aspect of entanglement, is not exclusive to quantum systems, and this principle applies to both, non-local and local entanglement, at the single- and multi-photon levels. The former is observed between photons that simultaneously exist in physically separated locations, and the latter between the internal degrees of freedom of photons. Classical states of light, such as complex vector light modes, also exhibit local entanglement, attributed to the non-separability between their spatial and polarisation degrees of freedom  \cite{Rosales2018Review,toninelli2019concepts}. Such non-separable modes are controversially called "classically-entangled" states, being this a subject of current debate  \cite{Karimi2015,konrad2019quantum,forbes2019classically}. Despite this controversy, it is becoming clear that entanglement can be studied with classically-entangled states providing an advantageous tool in several research fields \cite{Eberly2016,Qian2011,Qian2017}. For example, classical-entanglement has been exploited in quantum error correction \cite{Ndagano2017}, optical communications \cite{Ndagano2018,Milione2015,Wang2015} and optical metrology \cite{Hu2019,Toppel2014,BergJohansen2015}. Additionally, the tightly focusing properties of complex vector modes have been exploited in the field of optical tweezers \cite{Shvedov2014,Bhebhe2018a,Bhebhe2018,Kozawa2010,Skelton2013,Donato2012,Roxworthy2010},  micromachining \cite{Kraus2010}, as well as in super-resolution microscopy  \cite{Torok2004,Hao2010,Segawa2014}. 

In recent time, there has been an increasing interest in the engineering of vector light beams whose polarisation state varies upon free-space propagation. Most of these have focused on the generation of pure vector beams that oscillate from one vector state to another while keeping a constant degree of entanglement \cite{Moreno2015,ShiyaoFu2016,Davis2016,PengLi2017,PengLi2018,PengLi2016}. A more interesting case, which encloses the previous cases, reported the generation of light beams with a degree of entanglement oscillating between scalar and vector, offering a tool for the on-demand delivery of specific states to desired positions \cite{Otte2018}. Such oscillating modes are generated from the superposition of two counter-propagating vector beams, whose implementation can be cumbersome. While these studies have only considered cylindrical vector vortex modes, the use of new symmetries (spatial shapes), such as parabolic or elliptical, could potentially allow us to unveil properties of vector modes that up to now have remained hidden.

Here we demonstrate a new class of vector modes whose degree of entanglement decays as function of their propagation distance, evolving from a non--homogeneously polarized vector beam to a homogeneously polarized beam. Such modes are generated from a non-separable superposition of orthogonal parabolic beams, which are natural solutions to the Helmholtz equation in parabolic coordinates \cite{Bandres2004, Lopez-Mariscal2005, Bandres2008Accelerating, RodriguezLara2009, Ruelas2013, Gutierrez-Vega2005}, and orthogonal polarisation states. The entanglement decay is quantified through a modified measurement of concurrence $C$, commonly used to measure the degree of non-separability in vector modes \cite{McLaren2015,Selyem2019, Manthalkar2020, Zhao2020, Ndagano2016}.

\section{Theoretical Background} 
\begin{figure}[tb]
    \centering
    \includegraphics[width=0.48\textwidth]{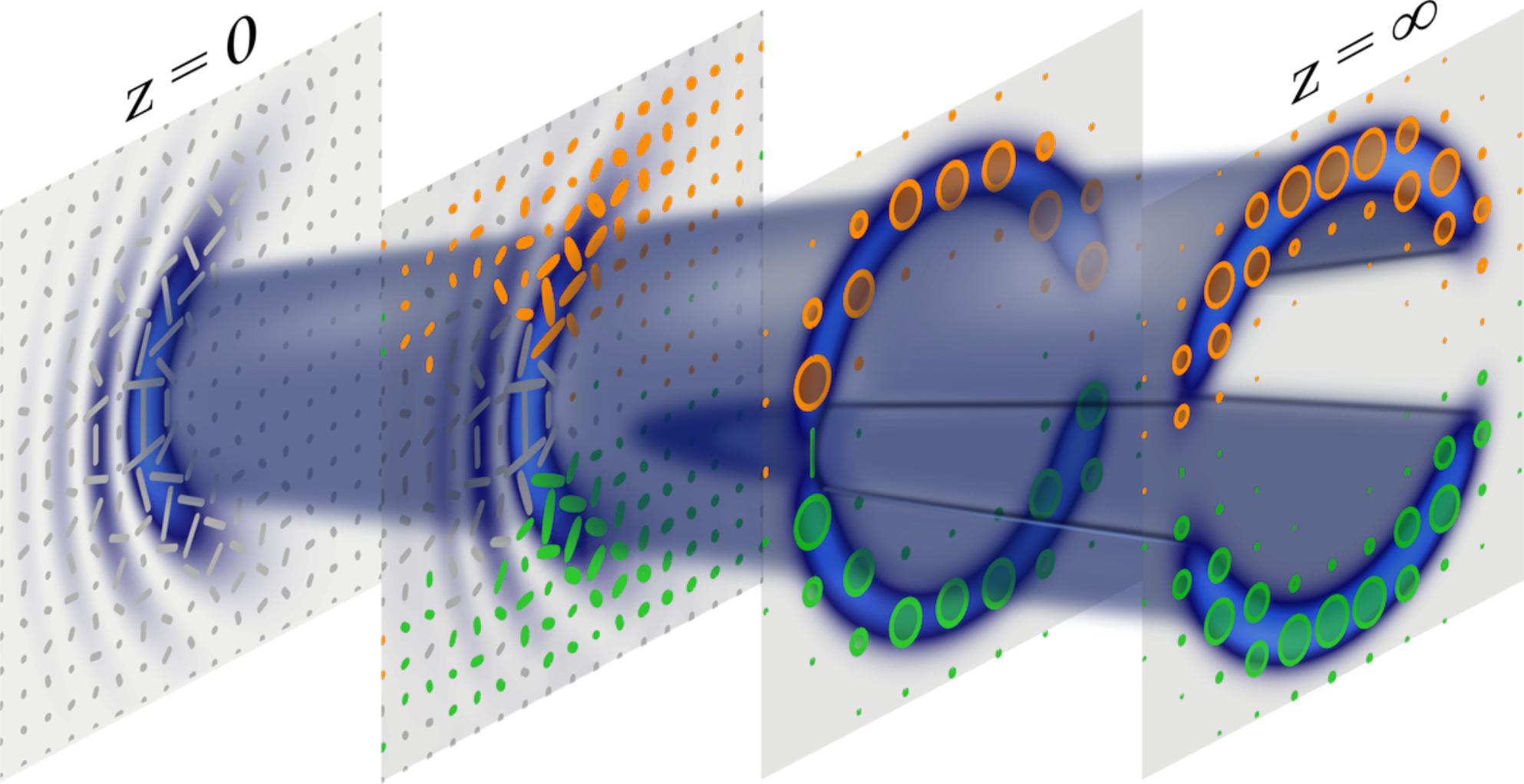}
    \caption{Schematic representation of a classically-entangled light mode featuring entanglement decay upon free space propagation. Right Circular Polarisation (RCP) is shown in orange, Left Circular Polarisation (LCP) in green and linear polarisation in gray.}
    \label{Concept}
\end{figure}

The concept presented in this manuscript is schematically shown in Fig. \ref{Concept}. At the generation plane ($z=0$), the parabolic vector beam possesses a maximum degree of entanglement, clearly evinced as a non-homogeneous polarisation distribution, overlapped with the intensity profile of the vector mode. Upon propagation, the polarisation structure of the beam evolves from completely mixed and locally non--seperable to completely unmixed and locally separable, the latter reached in the far field ($z=\infty$). 

The engineered vector modes are constructed from a superpositions of Traveling Parabolic-Gaussian ($\rm TPG^{\pm}$) beams. Mathematically, $\rm TPG$ beams are given by superposition of the even and odd Parabolic-Gaussian (PG) beams as \cite{Bandres2004}
\begin{equation}\label{TPGModes}
    \begin{split}
    \centering
    &\rm TPG^{\pm}({\bf r}; a)={\rm PG^{e}}({\bf r};a){\pm} i {\rm PG}^{o}({\bf r};a).
    \end{split}
\end{equation}
Here, the functions $\rm PG^{e,o}(\cdot)$ are the even and odd PG modes of the parabolic cylindrical coordinates ${\bf r}= (\eta,\xi,z)$ given by \cite{Gutierrez-Vega2005}
\begin{equation}\label{TPGModes}
    \begin{split}
    \centering
    {\rm PG^{e,o}}({\bf r};a)=&\exp\left( -i \frac{k_t^2}{2k}\frac{z}{\mu}\right){\rm GB}({\bf r})\frac{|\Gamma_1|^2}{\pi\sqrt{2}}{\rm P_{e,o}}\left( \sqrt{\frac{2k_t}{\mu}}\xi;a\right)\\
    &\times{\rm P_{e,o}}\left( \sqrt{\frac{2k_t}{\mu}}\eta;-a\right),
    \end{split}
\end{equation}
where $\rm P_e(\cdot)$ and $\rm P_o(\cdot)$ are the even and odd solutions of the parabolic cylindrical differential equation $[d^2/dx^2+(x^2/4-a)]{\rm P}(x;a)=0$ and $a \in (-\infty,\infty)$ represents the continuous order of the beam. Importantly, the solutions $\rm P_e(\cdot)$ and $\rm P_o(\cdot)$ can be written in terms of Kummer confluent hypergeometric functions \cite{NIST2010}, which are available in many numerical libraries. Furthermore, $\Gamma_1=\Gamma((1/4)+(1/2)ia)$, with  $\Gamma(\cdot)$ the gamma function, and $k_t$ is the transverse component of the wave vector, ${\bf k}$, whose magnitude is related to the longitudinal component $k_z$ as  $k^2=k_t^2+k_z^2$. Additionally, ${\rm GB}({\bf r})$ is the fundamental Gaussian beam given by
\begin{equation}
    {\rm GB}({\bf r})=\frac{\exp(ikz)}{\mu}\exp\left(-\frac{r^2}{\mu\omega_0^2}\right),
\end{equation}
where $\mu=\mu(z)=1+iz/z_r$ with $z_r=k\omega_0^2/2$ being the usual Rayleigh range of a Gaussian beam. The parabolic coordinates ${\bf r}= (\eta,\xi,z)$ are related to the Cartesian coordinates as $x=(\eta^2-\xi^2)/2$ and $y=\eta\xi$, where $\eta \in [0,\infty)$, $\xi \in (-\infty,\infty)$. For values $\gamma=k_t\omega_o\gg1$, the PG beam propagates in a non-diffracting way within the range $[-z_{max},z_{max}]$, where $z_{max}=\omega_0 k/k_t$.

The Travelling Parabolic-Gaussian Vector beams $\rm (TPGV)$ are generated as a non-separable weighted superposition of the polarisation and spatial degrees of freedom. Here, the polarisation degree of freedom is encoded in the circular polarization basis while the spatial is precisely encoded in the $\rm TPG^{\pm}(\cdot)$ modes of the parabolic cylindrical coordinates. Mathematically, such superposition can be written as
\begin{equation}
{\rm TPGV}({\bf r};a)=\frac{1}{\sqrt{2}}\left[{\rm TPG}^+({\bf r};a)\hat{e}_R+{\rm TPG}^-({\bf r};a)\exp(i\phi)\hat{e}_L\right],
\label{TPVModes}
\end{equation}
where the unitary vectors $\hat{e}_R$ and $\hat{e}_L$ represent the right and left circular states of polarisation, respectively. Finally, the term $\exp(i\phi)$ ($\phi\in[-\pi/4,\pi/4]$) is a phase difference between both constituting modes.

\begin{figure}[h!]
    \centering
    \includegraphics[width=0.45\textwidth]{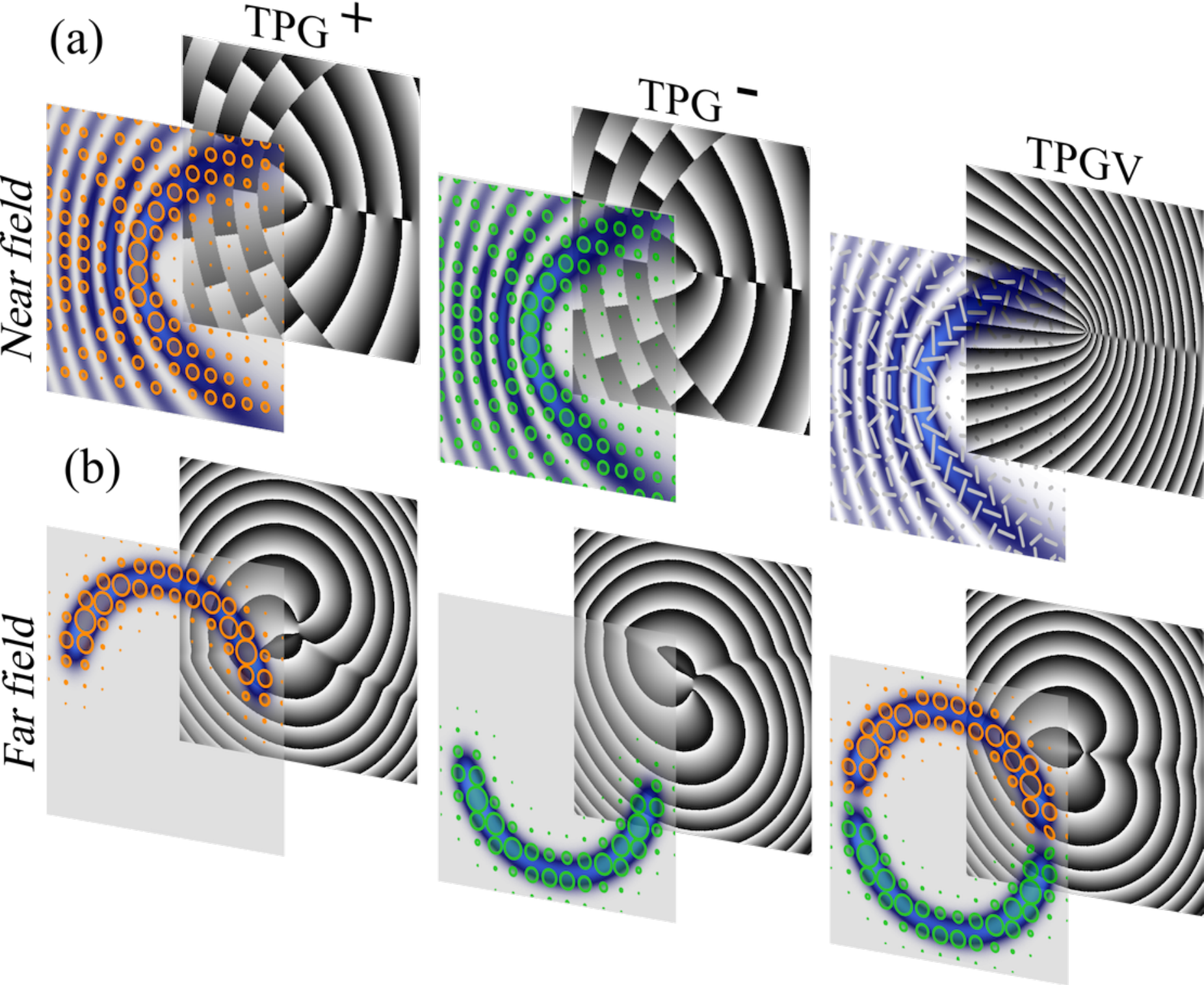}
    \caption{Phase (back panels) and polarisation distribution overlapped with the intensity profiles (front panels) of the scalar modes $\rm TPG^{+}({\bf r};a)\hat{e}_R$ (left) and $\rm TPG^{-}({\bf r};a)\hat{e}_L$, which are combined to generate the ${\rm TPGV}$ mode (right) at the (a) near- and (b) far-field. In the case of the TPGV mode, we depict the phase of the complex Stokes field $S=S_1+iS_2$}.
    \label{phase}
\end{figure}

An example of such a superposition is schematically illustrated in Fig. \ref{phase} using the modes ${\rm TPG}^{+}({\bf r}; 3)$ and ${\rm TPG}^{-}({\bf r}; 3)$, where the back panels show the phase distribution, whereas the front ones show the intensity profile overlapped with the corresponding polarisation distribution. In Fig.~\ref{phase}(a), we present the near-field of each individual component (left and middle panels) as well as the complex superposition (right panel), and in Fig.~\ref{phase}(b) their corresponding far-field counterparts. Notice the difference between the near-field, a completely mixed and locally non--separable, and the far--field, completely unmixed and locally separable.

\section{Experimental generation of TPGV modes}

\begin{figure*}[tb]
    \centering
    \includegraphics[width=0.98\textwidth]{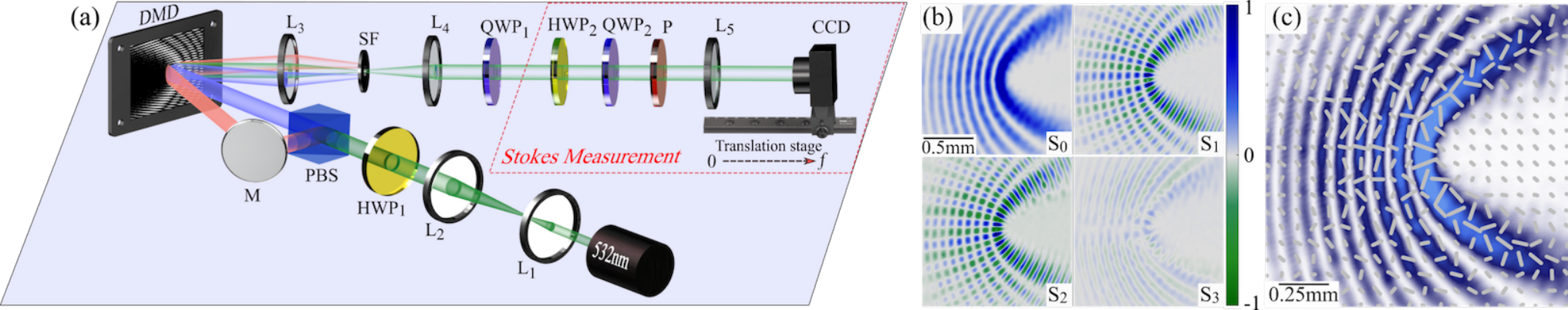}
    \caption{(a) To experimentally generate TPGV mode, we used a novel approach based on a Digital Micromirror Devices (DMD). An expanded and collimated ( by lenses L$_1$ and L$_2$ ) laser beam is diagonally polarised with a Half-wave plate (HWP$_1$) and split by a Polarising beam splitter (PBS) according to its polarisation components. The two beams are redirected to the DMD impinging at slightly different angles but overlapped at the center of a binary multiplexed hologram where the $\rm TPG^{+}$ and  $\rm TPG^{-}$ scalar modes are encoded, each with a unique linear grating. After the DMD, the first diffraction order of each beam overlaps along a common propagation axis where the TPGV is generated. A spatial filter (SF) placed at the focusing plane of a telescope formed by lenses L$_3$ and L$_4$ removes all higher diffraction orders. A quarter wave plate (QWP$_1$) transforms the mode from the linear to the circular polarisation basis. The entanglement decay is quantified through Stokes poalrimetry, for which a set of four intensities are recorded with a Charge-Coupled Device camera (CCD). (b) Experimental Stokes parameters  $S_1$, $S_2$ and $S_3$ of the TPGV mode $\rm{TPGV}({\bf r},3)$. (c) Intensity profile overlapped with the reconstructed polarisation distribution.}
    \label{setup}
\end{figure*}
A schematic representation of our experimental setup to generate arbitrary vector modes is depicted in Fig.\ref{setup}(a). A horizontally polarised laser beam ($\lambda=532$ nm) is expanded and collimated by lenses L$_1$ and L$_2$, and subsequently transformed into a diagonally polarised beam by use of a Half-wave plate (HWP) at $45^\circ$. A Polarising Beam Splitter (PBS) separates the beam into its horizontal and vertical polarization components. Both beams are then directed, one with the help of a mirror (M), to a polarisation-independent Digital Micromirror Device (DMD, DLP Light Crafter 6500 from Texas Instruments), impinging under slightly different angles ($\approx 1.5^\circ$) at the center of the DMD. Here, a multiplexed hologram consisting of the superposition of two independent holograms with unique linear phase gratings, each corresponding to the constituting wave fields of Eq. \ref{TPVModes}, ${\rm TPG}^+$ and ${\rm TPG}^-$ modes. In this way, the period of the grating is used to ensure the first diffraction order of each beam to propagate along a common axis, where the vector beam is generated (see \cite{Rosales2020} for further details). Once generated, a spatial filter (SF) placed at the focusing point of a telescope composed by lenses $\rm L_3$ and $\rm L_4$ removes all higher diffraction orders. A Quarter-Wave plate (QWP$_1$) is added to change the $\rm TPG$ mode from the linear ($\hat{e}_H,\hat{e}_V$) to the circular polarisation basis ($\hat{e}_L,\hat{e}_R$). In order to reach the far-field, a lens $\rm L_5$ of focal length $f=300$ mm was inserted in the path of the beam. A Charge-coupled device camera (CCD) mounted on a rail parallel to the propagation direction of the beam was used to record its intensity. 

To quantify the degree of entanglement we relied on a well--known measure from quantum mechanics, the concurrence $C$, which assigns a value in the rage $[0,1]$ to the degree of entanglement \cite{McLaren2015, Ndagano2016, Selyem2019, Manthalkar2020, Zhao2020}. The concurrence $C$ is measured by integrating the Stokes parameters $S_i, i=0,1,2,3$  over the entire transverse plane through the relation \cite{Selyem2019,Manthalkar2020},
\begin{equation}
C=\sqrt{1-\left(\frac{\mathbb{S}_1}{\mathbb{S}_0} \right)^2-\left(\frac{\mathbb{S}_2}{\mathbb{S}_0} \right)^2-\left(\frac{\mathbb{S}_3}{\mathbb{S}_0} \right)^2},
\label{concurrence}
\end{equation}
where $\mathbb{S}_i = \iint_{R^2} S_i \:dA$.  The Stokes parameters $S_i$ are computed from a set of four intensity measurements as \cite{Goldstein2011}
\begin{equation}\label{Eq.Stokes}
\begin{split}
\centering
  &S_{0}=I_{0},\hspace{19mm} S_{1}=2I_{H}-S_{0},\hspace{1mm}\\
 &S_{2}=2I_{D}-S_{0},\hspace{10mm} S_{3}=2I_{R}-S_{0},
\end{split}
\end{equation}
where $I_0$ is the total intensity of the mode and $I_H$, $I_D$ and $I_R$ the intensity of the horizontal, diagonal and right-handed polarisation components, respectively. As illustrated in Fig. \ref{setup}(a), such intensity measurements were acquired by a CCD camera through the combination of a linear polariser (P), and a Quarter-wave plate (QWP$_2$) \cite{Zhao2019}. Specifically, the intensities of the horizontal ($I_H$) and diagonal ($I_D$) polarisation components, were obtained by passing the beam through a linear polariser at $0^\circ$ and $45^\circ$, respectively, whereas intensity corresponding to the RCP component ($I_R$) by passing the beam simultaneously through a QWP at $45^\circ$ and a linear polariser at $90^\circ$. As a way of example, Fig. \ref{setup}(b) shows the experimental Stokes parameters $S_{0}$, $S_{1}$, $S_{2}$ and $S_{3}$ for the specific mode $\rm{TPGV}({\bf r},3)$ at $z=0$. Such parameters were used to reconstruct the transverse polarisation distribution on a $20\times20$ grid, as shown in Fig. \ref{setup}(c). Here, for the sake of clarity, we also display the transverse intensity profile. For this specific example, we have $\mathbb{S}_1/\mathbb{S}_0=0.12$, $\mathbb{S}_2/\mathbb{S}_0=0.09$ and $\mathbb{S}_3/\mathbb{S}_0=-0.02$ that upon substitution in Eq. \ref{concurrence} yield the value $C=0.98$, which, as expected, corresponds to a maximally entangled mode.  

\section{Free-space local entanglement decay analysis}
\begin{figure}[tb]
    \centering
    \includegraphics[width=0.48\textwidth]{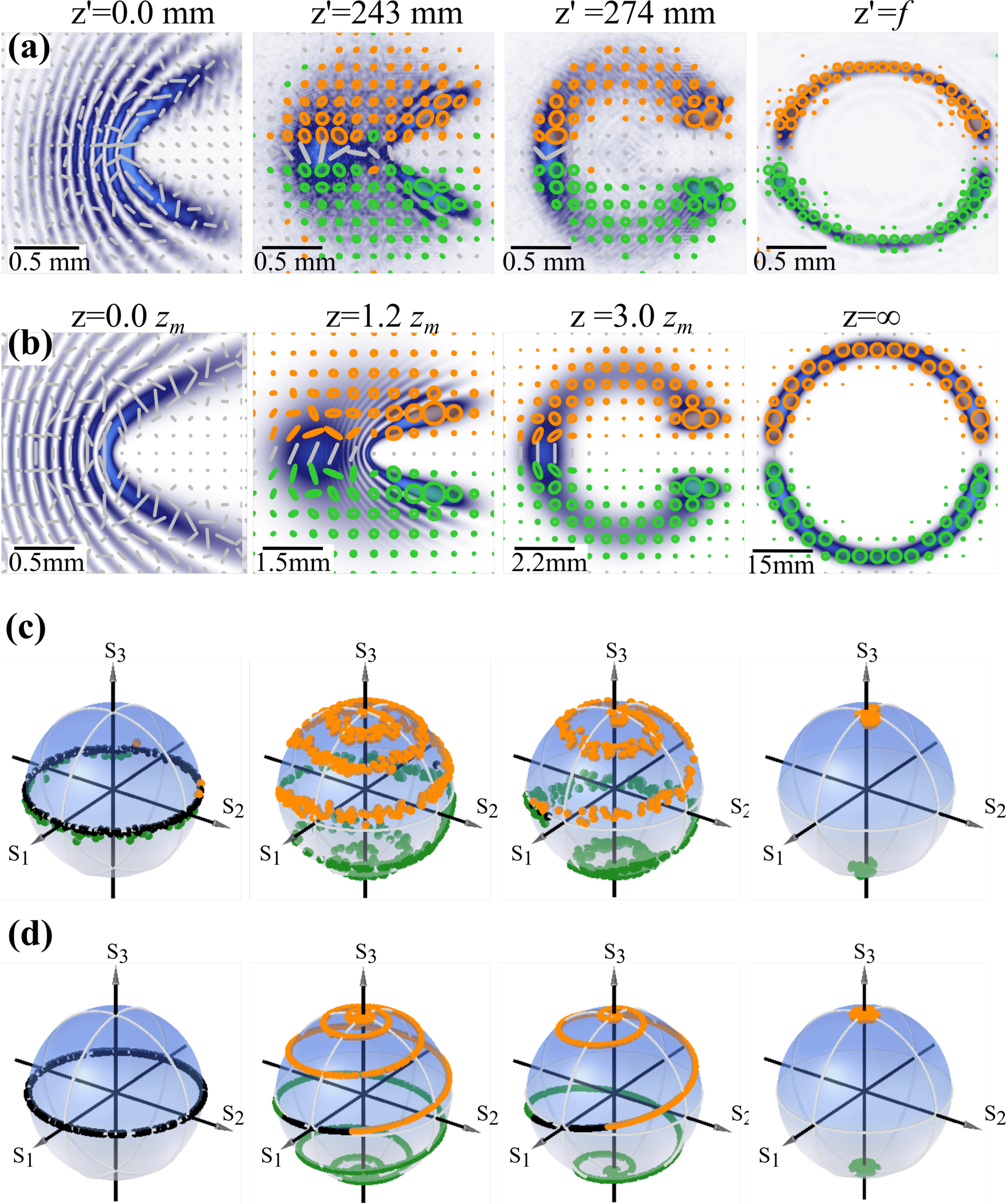}
    \caption{Experimental (a) and simulated (b) evolution of intensity and polarisation distribution of TPGV beams as function of propagation distance. Experimental (c) and (d) simulated representation of the transverse polarisation distribution on a Poincar\'e sphere, each in correspondence with the planes shown in (a) and (b).}
    \label{Polarisation}
\end{figure}
In this section, we present a detailed analysis of the propagation dynamics of our TPGV beams. This analysis is performed experimentally and through numerical simulations using the Rayleigh–Sommerfeld diffraction theory \cite{Goodman1996}. As a first evidence of the evolution dynamics, we reconstructed the polarisation distribution at various transverse planes. A representative set of results are displayed in Fig. \ref{Polarisation} (a) for the case $ {\rm TPGV}({\bf r};3)$ for $\omega_0=2$ mm and $k_t=22.5$ mm$^{-1}$, for which, $z_{max}=1050$ mm. Here, it is shown the transverse intensity profile overlapped with its corresponding polarisation distribution at four different planes, namely $z=0$, $z=1.2 z_{max}$, $z=3.0 z_{max}$ and $z=\infty$. Experimentally, these distances correspond to the values $z'=243$ mm, $z'=274$ mm and $z'=f$, obtained through the well-known equation of a thin lens, $1/z+1/z'=1/f$. As shown, the plane $z=0$ features a vector mode with a non-homogeneous polarisation distribution. For $z>0$, the beam evolves from completely mixed and locally non--separable to completely unmixed and locally separable, featuring a smooth transition from linear to circular polarisation. For comparison, in Fig. \ref{Polarisation}(b) we show the theoretical counterparts, where the same behaviour is observed.

With the idea of better visualising the evolution of polarisation upon propagation, we mapped the different states of polarisation acquired at each plane onto the well-known Poincar\'e sphere (PS), in which the different polarisation states are associated to unique points on the surface of the sphere \cite{Goldstein2011}, as shown in Fig. \ref{Polarisation}(c) and \ref{Polarisation}(d) for experiment and numerical simulation, respectively. Here, the coordinate axis are given in terms of the Stokes parameters $S_1$, $S_2$ and $S_3$. For $z=0$, all the states of polarisation in the transverse plane are linear and therefore mapped to points along the equator. For $z>0$, such linear polarisation states gradually evolve from linear into elliptical and finally to circular. In the Poincar\'e Sphere, this is seen as points along a spiral trajectory connecting the North and South poles, in the intermediate planes, and in the far-field as points on the North and South poles. Notice that even though the polarisation structure evolves from completely mixed to completely unmixed, the amount of right- and left-elliptically polarised photons remains in equilibrium.

\begin{figure}[tb]
    \centering
    \includegraphics[width=.490\textwidth]{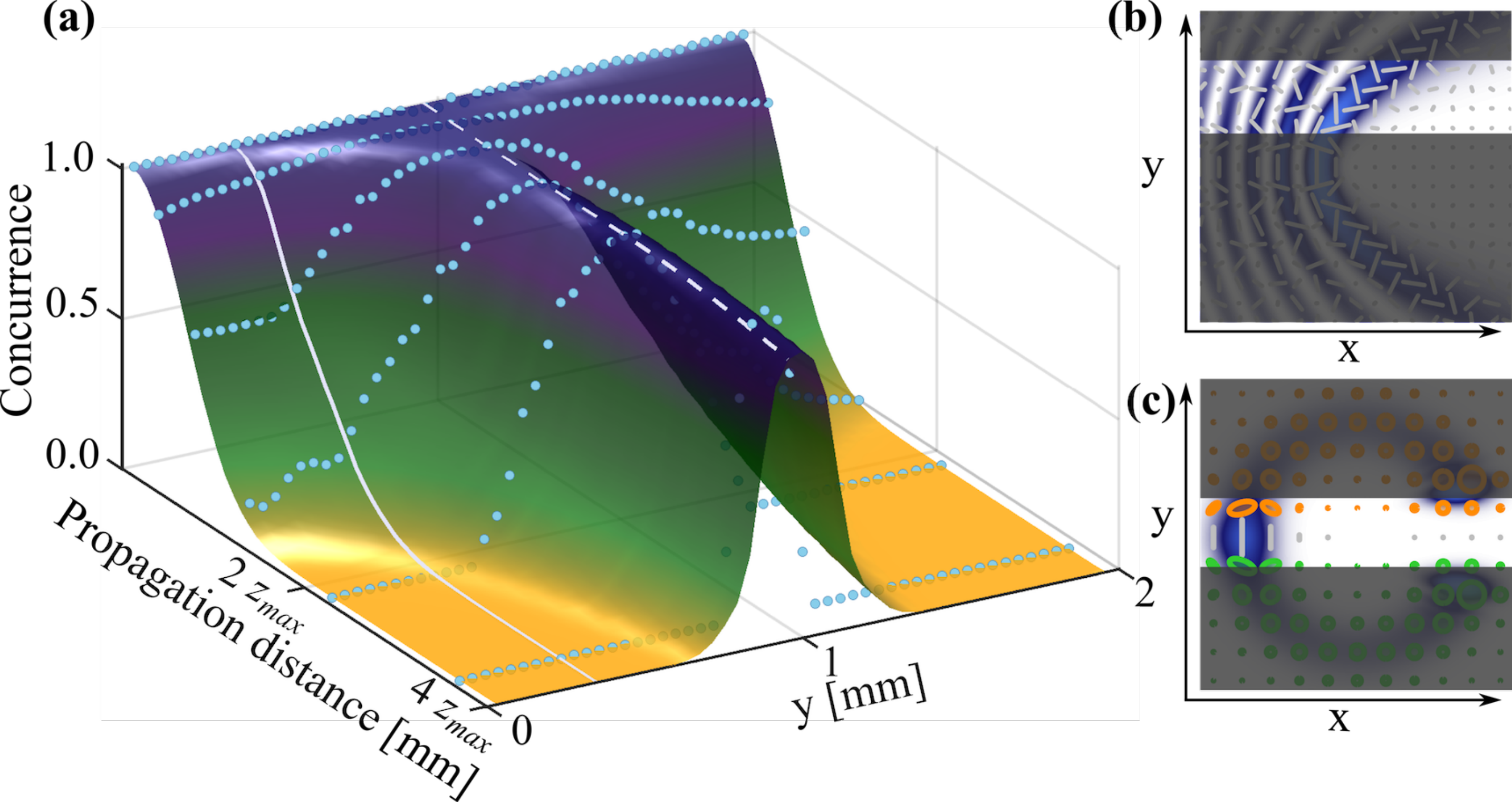}
    \caption{(a) Concurrence as function of propagation and transverse coordinates. The solid and dashed lines show two examples of $C$ as function of $z$ for the specific sections of the beam, shown in (b) and (c), which correspond to $z=0$ and $z=4 z_{max}$, respectively. The surface corresponds to numerical simulations and the data points to experimental measurements at the propagation planes $z/z_{max}=0, 0.2, 0.7, 1.3, 2.2$ and $3.6$.  An analogous behaviour has been reported in the context of temporal coherence \cite{Yang2015}, where the generation of beams whose local degree of temporal coherence varies as a function of the time difference is discussed.}
    \label{prop1}
\end{figure}

One way to quantify the entanglement decay of the TPGV modes, as function of the propagation distance $z$ (Eq. \ref{concurrence}), is through the computation of the concurrence, $C$. Nonetheless, the concurrence of the whole beam, that we call global concurrence, yields $C\approx 1$ for every propagation plane. The reason being, as mentioned before, the amount of left- and right-elliptically polarised photons is always the same regardless of the propagation distance. Hence, even though locally all the Stokes parameters might have non zero values, globally $S_1$, $S_2$ and $S_3$ will be always zero, resulting in $\mathbb{S}_i = \iint_{R^2} S_i \:dA=0$, for $i=1,2,3$, which implies $C=1$ for all values of $z$. This implies that the global concurrence fails to account for the changes in the polarisation distribution that are observed in Fig.~\ref{Polarisation}. This can be illustrated by computing a local concurrence $C_L$ , {\it i.e.}, the concurrence of small sections of the beam, across the entire transverse plane and at different propagation distances. This is shown in Fig. \ref{prop1} (a), where we performed a numerical simulation of $C_L$ along the vertical direction across the beam and as function of the propagation distance. Here, for each $z\in[0,4 z_{max}]$, we sectioned the beam from top to bottom into 50 rectangular regions (2 mm $\times$ 40 $\mu$m each) and computed the concurrence on each of them. As can be seen, for short propagation distances, $z\approx0$, the concurrence stays closer to one for all beam positions, which is expected since at the origin plane both degrees of freedom are completely mixed over the full transverse plane. Figure \ref{prop1} (b) shows an example of the area over which $C$ was computed for the specific case $z=0$. This area corresponds to the solid line shown in Fig. \ref{prop1}, which illustrates a decay of $C$ as function of $z$. Notice that in the centre of the beam $C$ remains constant regardless of the propagation distance. To further understand this, in Fig.~\ref{prop1} (c) we show the integration area for this particular position at $z=3 z_{max}$, which clearly shows that even though in this region both polarisation states are completely separated, it contains the same amount of right- and left- elliptically polarised photons, yielding $C=1$. The corresponding concurrence  as function of the propagation area for this area is shown as a dashed line in Fig. \ref{prop1}. In addition to the simulation, in this figure we also present experimental values of the concurrence for the cases $z/z_{max}=0, 0.2, 0.7, 1.3, 2.2$ and $3.6$, shown as points over the 3D surface, which agree very well with the simulation. Figure ~\ref{prop1} clearly evinces that local concurrence is the appropriate measure to quantify the decay in degree of local entanglement, even though it fails in the centre of the beam, where both polarisation components are present.  Some studies have explored the distribution of polarisation states on the Poincar\'e sphere \cite{Refregier2005, kagalwala2013bell}, however, they do not address their spatial distribution across the transverse plane of the beam. To this end, we envision the definition of a new parameter to quantify the inhomogeneous distribution of polarization states for vector beams. However, in this work it suffices to stress out that local entanglement decay can be observed from the computation of the local concurrence.  

\section{Conclusions}
In this work we demonstrated a novel kind of complex light field that upon free-space propagation evolves from maximally mixed and locally non--separable to completely unmixed and locally separable. More precisely, we generated a vector beam with a non-homogeneous polarisation distribution that upon free-space propagation evolves into a homogeneously polarized mode.  Such behaviour is directly observed at various propagation distances, through a reconstruction of the transverse polarisation distribution performed via Stokes polarimetry. This is further evinced by mapping the entire polarisation distribution at each plane onto the Poincar\'e sphere, which exhibits an evolution of the state of polarisation from the equator to the poles. Such evolution happens along spirals connecting the North and South poles. A quantification of such local entanglement decay was performed through the concurrence $C$, which takes the value $C=1$, for vector states, and $C=0$ for scalar beams. We noted that a global measure of concurrence can not properly quantify the entanglement decay upon propagation even though this is evident from the transverse polarisation distribution reconstructed at various propagation planes. Nonetheless, a ``local concurrence'', computed in smaller sections of the beam, clearly shows that $C$ decreases as function of the propagation distance, reaching the value $C=0$ in the far field. This evinces the need for an alternative definition of $C$, which takes into account the local variations of the non-separability but it lies beyond the scope of this research. Importantly, the entanglement decay reported here can not be obtained with cylindrical vector beams as it is an intrinsic property of parabolic vector beams. It is also worth mentioning that it can be adjusted through the parameter $z_{max}$, which depends on $\omega_0$ and $k_t$. Finally, these novel states of light offer a new tool for a wide variety of applications in fields such as optical communications, optical metrology and optical tweezers, to mention a few. 

\section*{Funding}
This work was partially supported by the National Natural Science Foundation of China (NSFC) under Grant No.  61975047.  BPG and RIHA acknowledge support from Consejo Nacional de Ciencia y Tecnolog\'ia (PN2016-3140).

\section*{Disclosures}
The authors declare that there are no conflicts of interest related to this article.
\bibliography{References}

\begin{thebibliography}{10}
\newcommand{\enquote}[1]{``#1''}

\bibitem{Rosales2018Review}
C.~Rosales-Guzm\'{a}n, B.~Ndagano, and A.~Forbes, \enquote{{A review of complex
  vector light fields and their applications},} {\protect\JournalTitle{J.
  Opt.}} \textbf{20}, 123001 (2018).

\bibitem{toninelli2019concepts}
E.~Toninelli, B.~Ndagano, A.~Vall{\'e}s, B.~Sephton, I.~Nape, A.~Ambrosio,
  F.~Capasso, M.~J. Padgett, and A.~Forbes, \enquote{Concepts in quantum state
  tomography and classical implementation with intense light: a tutorial,}
  {\protect\JournalTitle{Advances in Optics and Photonics}} \textbf{11},
  67--134 (2019).

\bibitem{Karimi2015}
E.~Karimi and R.~W. Boyd, \enquote{{Classical entanglement?}}
  {\protect\JournalTitle{Science}} \textbf{350}, 1172--1173 (2015).

\bibitem{konrad2019quantum}
T.~Konrad and A.~Forbes, \enquote{Quantum mechanics and classical light,}
  {\protect\JournalTitle{Contemporary Physics}} pp. 1--22 (2019).

\bibitem{forbes2019classically}
A.~Forbes, A.~Aiello, and B.~Ndagano, \enquote{Classically entangled light,} in
  \emph{Progress in Optics,}  (Elsevier Ltd., 2019), pp. 99--153.

\bibitem{Eberly2016}
J.~H. Eberly, X.-F. Qian, A.~A. Qasimi, H.~Ali, M.~A. Alonso,
  R.~Guti{\'e}rrez-Cuevas, B.~J. Little, J.~C. Howell, T.~Malhotra, and A.~N.
  Vamivakas, \enquote{Quantum and classical optics--emerging links,}
  {\protect\JournalTitle{Physica Scripta}} \textbf{91}, 063003 (2016).

\bibitem{Qian2011}
X.-F. Qian and J.~H. Eberly, \enquote{Entanglement and classical polarization
  states,} {\protect\JournalTitle{Opt. Lett.}} \textbf{36}, 4110--4112 (2011).

\bibitem{Qian2017}
X.-F. Qian, A.~N. Vamivakas, and J.~H. Eberly, \enquote{Emerging connections:
  Classical and quantum optics,} {\protect\JournalTitle{Opt. Photon. News}}
  \textbf{28}, 34--41 (2017).

\bibitem{Ndagano2017}
B.~Ndagano, B.~Perez-Garcia, F.~S. Roux, M.~McLaren, C.~Rosales-Guzm\'{a}n,
  Y.~Zhang, O.~Mouane, R.~I. Hernandez-Aranda, T.~Konrad, and A.~Forbes,
  \enquote{{Characterizing quantum channels with non-separable states of
  classical light},} {\protect\JournalTitle{Nature Phys.}} \textbf{13},
  397--402 (2017).

\bibitem{Ndagano2018}
B.~Ndagano, I.~Nape, M.~A. Cox, C.~Rosales-Guzm\'{a}n, and A.~Forbes,
  \enquote{Creation and detection of vector vortex modes for classical and
  quantum communication,} {\protect\JournalTitle{J. Light. Technol.}}
  \textbf{36}, 292--301 (2018).

\bibitem{Milione2015}
G.~Milione, M.~P. Lavery, H.~Huang, Y.~Ren, G.~Xie, T.~A. Nguyen, E.~Karimi,
  L.~Marrucci, D.~A. Nolan, R.~R. Alfano \emph{et~al.}, \enquote{{4$\times$20
  Gbit/s mode division multiplexing over free space using vector modes and a
  q-plate mode (de) multiplexer},} {\protect\JournalTitle{Opt. Lett.}}
  \textbf{40}, 1980--1983 (2015).

\bibitem{Wang2015}
Y.~Zhao and J.~Wang, \enquote{{High-base vector beam encoding/decoding for
  visible-light communications},} {\protect\JournalTitle{Opt. Lett.}}
  \textbf{40}, 4843--4846 (2015).

\bibitem{Hu2019}
X.-B. Hu, B.~Zhao, Z.-H. Zhu, W.~Gao, and C.~Rosales-Guzm\'{a}n, \enquote{In
  situ detection of a cooperative target's longitudinal and angular speed using
  structured light,} {\protect\JournalTitle{Opt. Lett.}} \textbf{44},
  3070--3073 (2019).

\bibitem{Toppel2014}
F.~T{\"o}ppel, A.~Aiello, C.~Marquardt, E.~Giacobino, and G.~Leuchs,
  \enquote{Classical entanglement in polarization metrology,}
  {\protect\JournalTitle{New J. Phys.}} \textbf{16}, 073019 (2014).

\bibitem{BergJohansen2015}
S.~Berg-Johansen, F.~T\"{o}ppel, B.~Stiller, P.~Banzer, M.~Ornigotti,
  E.~Giacobino, G.~Leuchs, A.~Aiello, and C.~Marquardt, \enquote{Classically
  entangled optical beams for high-speed kinematic sensing,}
  {\protect\JournalTitle{Optica}} \textbf{2}, 864--868 (2015).

\bibitem{Shvedov2014}
V.~Shvedov, A.~R. Davoyan, C.~Hnatovsky, N.~Engheta, and W.~Krolikowski,
  \enquote{A long-range polarization-controlled optical tractor beam,}
  {\protect\JournalTitle{Nature Photonics}} \textbf{8}, 846--850 (2014).

\bibitem{Bhebhe2018a}
N.~Bhebhe, C.~Rosales-Guzman, and A.~Forbes, \enquote{Classical and quantum
  analysis of propagation invariant vector flat-top beams,}
  {\protect\JournalTitle{Appl. Opt.}} \textbf{57}, 5451--5458 (2018).

\bibitem{Bhebhe2018}
N.~Bhebhe, P.~A.~C. Williams, C.~Rosales-Guzm{\'a}n, V.~Rodriguez-Fajardo, and
  A.~Forbes, \enquote{A vector holographic optical trap,}
  {\protect\JournalTitle{Sci. Rep.}} \textbf{8}, 17387 (2018).

\bibitem{Kozawa2010}
Y.~Kozawa and S.~Sato, \enquote{{Optical trapping of micrometer-sized
  dielectric particles by cylindrical vector beams},}
  {\protect\JournalTitle{Opt. Express}} \textbf{18}, 10828--10833 (2010).

\bibitem{Skelton2013}
S.~E. Skelton, M.~Sergides, R.~Saija, M.~a. Iat{\`{i}}, O.~M. Marag{\'{o}}, and
  P.~H. Jones, \enquote{{Trapping volume control in optical tweezers using
  cylindrical vector beams.}} {\protect\JournalTitle{Opt. Lett.}} \textbf{38},
  28--30 (2013).

\bibitem{Donato2012}
M.~G. Donato, S.~Vasi, R.~Sayed, P.~H. Jones, F.~Bonaccorso, A.~C. Ferrari,
  P.~G. Gucciardi, and O.~M. Marag\'{o}, \enquote{{Optical trapping of
  nanotubes with cylindrical vector beams},} {\protect\JournalTitle{Opt.
  Lett.}} \textbf{37}, 3381 (2012).

\bibitem{Roxworthy2010}
B.~Roxworthy and K.~Toussaint, \enquote{{Optical trapping with $\pi$-phase
  cylindrical vector beams},} {\protect\JournalTitle{New J. Phys.}}
  \textbf{12}, 073012 (2010).

\bibitem{Kraus2010}
M.~Kraus, M.~A. Ahmed, A.~Michalowski, A.~Voss, R.~Weber, and T.~Graf,
  \enquote{Microdrilling in steel using ultrashort pulsed laser beams with
  radial and azimuthal polarization,} {\protect\JournalTitle{Opt. Express}}
  \textbf{18}, 22305--22313 (2010).

\bibitem{Torok2004}
P.~T{\"{o}}r{\"{o}}k and P.~Munro, \enquote{{The use of Gauss-Laguerre vector
  beams in STED microscopy},} {\protect\JournalTitle{Opt. Express}}
  \textbf{12}, 3605--3617 (2004).

\bibitem{Hao2010}
X.~Hao, C.~Kuang, T.~Wang, and X.~Liu, \enquote{{Effects of polarization on the
  de-excitation dark focal spot in STED microscopy},} {\protect\JournalTitle{J.
  Opt.}} \textbf{12}, 115707 (2010).

\bibitem{Segawa2014}
S.~Segawa, Y.~Kozawa, and S.~Sato, \enquote{{Resolution enhancement of confocal
  microscopy by subtraction method with vector beams.}}
  {\protect\JournalTitle{Opt. Lett.}} \textbf{39}, 3118--21 (2014).

\bibitem{Moreno2015}
I.~Moreno, J.~A. Davis, M.~M. S\'{a}nchez-L\'{o}pez, K.~Badham, and D.~M.
  Cottrell, \enquote{Nondiffracting {B}essel beams with polarization state that
  varies with propagation distance,} {\protect\JournalTitle{Opt. Lett.}}
  \textbf{40}, 5451--5454 (2015).

\bibitem{ShiyaoFu2016}
S.~Fu, S.~Zhang, and C.~Gao, \enquote{{B}essel beams with spatial oscillating
  polarization,} {\protect\JournalTitle{Scientific Reports}} \textbf{6}, 30765
  (2016).

\bibitem{Davis2016}
J.~A. Davis, I.~Moreno, K.~Badham, M.~M. S\'{a}nchez-L\'{o}pez, and D.~M.
  Cottrell, \enquote{Nondiffracting vector beams where the charge and the
  polarization state vary with propagation distance,}
  {\protect\JournalTitle{Opt. Lett.}} \textbf{41}, 2270--2273 (2016).

\bibitem{PengLi2017}
P.~Li, Y.~Zhang, S.~Liu, H.~Cheng, L.~Han, D.~Wu, and J.~Zhao,
  \enquote{Generation and self-healing of vector {B}essel-{G}auss beams with
  variant state of polarizations upon propagation,} {\protect\JournalTitle{Opt.
  Express}} \textbf{25}, 5821--5831 (2017).

\bibitem{PengLi2018}
P.~Li, D.~Wu, Y.~Zhang, S.~Liu, Y.~Li, S.~Qi, and J.~Zhao,
  \enquote{Polarization oscillating beams constructed by copropagating optical
  frozen waves,} {\protect\JournalTitle{Photon. Res.}} \textbf{6}, 756--761
  (2018).

\bibitem{PengLi2016}
P.~Li, Y.~Zhang, S.~Liu, L.~Han, H.~Cheng, F.~Yu, and J.~Zhao,
  \enquote{Quasi-{B}essel beams with longitudinally varying polarization state
  generated by employing spectrum engineering,} {\protect\JournalTitle{Opt.
  Lett.}} \textbf{41}, 4811--4814 (2016).

\bibitem{Otte2018}
E.~Otte, C.~Rosales-Guzm\'{a}n, B.~Ndagano, C.~Denz, and A.~Forbes,
  \enquote{Entanglement beating in free space through spin-orbit coupling,}
  {\protect\JournalTitle{Light: Science \& Applications}} \textbf{7}, e18009
  (2018).

\bibitem{Bandres2004}
M.~A. Bandres, J.~C. Guti\'{e}rrez-Vega, and S.~Ch\'{a}vez-Cerda,
  \enquote{Parabolic nondiffracting optical wave fields,}
  {\protect\JournalTitle{Opt. Lett.}} \textbf{29}, 44--46 (2004).

\bibitem{Lopez-Mariscal2005}
C.~L\'{o}pez-Mariscal, M.~A. Bandres, J.~C. Guti\'{e}rrez-Vega, and
  S.~Ch\'{a}vez-Cerda, \enquote{Observation of parabolic nondiffracting optical
  fields,} {\protect\JournalTitle{Opt. Express}} \textbf{13}, 2364--2369
  (2005).

\bibitem{Bandres2008Accelerating}
M.~A. Bandres, \enquote{Accelerating parabolic beams,}
  {\protect\JournalTitle{Opt. Lett.}} \textbf{33}, 1678--1680 (2008).

\bibitem{RodriguezLara2009}
B.~M. Rodr\'{\i}guez-Lara and R.~J\'auregui, \enquote{Dynamical constants of
  structured photons with parabolic-cylindrical symmetry,}
  {\protect\JournalTitle{Phys. Rev. A}} \textbf{79}, 055806 (2009).

\bibitem{Ruelas2013}
A.~Ruelas, S.~Lopez-Aguayo, and J.~C. Guti\'{e}rrez-Vega, \enquote{Engineering
  parabolic beams with dynamic intensity profiles,} {\protect\JournalTitle{J.
  Opt. Soc. Am. A}} \textbf{30}, 1476--1483 (2013).

\bibitem{Gutierrez-Vega2005}
J.~C. Guti\'{e}rrez-Vega and M.~A. Bandres, \enquote{Helmholtz--{G}auss waves,}
  {\protect\JournalTitle{J. Opt. Soc. Am. A}} \textbf{22}, 289--298 (2005).

\bibitem{McLaren2015}
M.~McLaren, T.~Konrad, and A.~Forbes, \enquote{{Measuring the nonseparability
  of vector vortex beams},} {\protect\JournalTitle{Phys. Rev. A}} \textbf{92},
  023833 (2015).

\bibitem{Selyem2019}
A.~Selyem, C.~Rosales-Guzm\'an, S.~Croke, A.~Forbes, and S.~Franke-Arnold,
  \enquote{Basis-independent tomography and nonseparability witnesses of pure
  complex vectorial light fields by {Stokes} projections,}
  {\protect\JournalTitle{Phys. Rev. A}} \textbf{100}, 063842 (2019).

\bibitem{Manthalkar2020}
A.~Manthalkar, I.~Nape, N.~T. Bordbar, C.~Rosales-Guzm\'{a}n, S.~Bhattacharya,
  A.~Forbes, and A.~Dudley, \enquote{All-digital stokes polarimetry with a
  digital micromirror device,} {\protect\JournalTitle{Opt. Lett.}} \textbf{45},
  2319--2322 (2020).

\bibitem{Zhao2020}
B.~Zhao, X.-B. Hu, V.~Rodríguez-Fajardo, A.~Forbes, W.~Gao, Z.-H. Zhu, and
  C.~Rosales-Guzmán, \enquote{Determining the non-separability of vector modes
  with digital micromirror devices,} {\protect\JournalTitle{Applied Physics
  Letters}} \textbf{116}, 091101 (2020).

\bibitem{Ndagano2016}
B.~Ndagano, H.~Sroor, M.~McLaren, C.~Rosales-Guzm{\'{a}}n, and A.~Forbes,
  \enquote{{Beam quality measure for vector beams},}
  {\protect\JournalTitle{Opt. Lett.}} \textbf{41}, 3407 (2016).

\bibitem{NIST2010}
F.~Olver, N.~I. of~Standards, T.~(U.S.), D.~Lozier, R.~Boisvert, and C.~Clark,
  \emph{NIST Handbook of Mathematical Functions Hardback and CD-ROM} (Cambridge
  University Press, 2010).

\bibitem{Rosales2020}
C.~Rosales-Guzm\'an, H.~Xiao-Bo, A.~Selyem, P.~Moreno-Acosta, F.-A. Sonja,
  R.~Ramos-Garcia, A.~Forbes, and S.~Franke-Arnold,
  \enquote{Polarisation-insensitive generation of vector modes using a digital
  micromirror device,} {\protect\JournalTitle{arXiv:2002.07843}}  (2020).

\bibitem{Goldstein2011}
D.~H. Goldstein, \emph{Polarized light} (CRC Press, 2011).

\bibitem{Zhao2019}
B.~Zhao, X.-B. Hu, V.~Rodr\'{i}guez-Fajardo, Z.-H. Zhu, W.~Gao, A.~Forbes, and
  C.~Rosales-Guzm\'{a}n, \enquote{Real-time {Stokes} polarimetry using a
  digital micromirror device,} {\protect\JournalTitle{Opt. Express}}
  \textbf{27}, 31087--31093 (2019).

\bibitem{Goodman1996}
J.~W. Goodman, \emph{Introduction to Fourier Optics} (WH Freeman, New York,
  2017), 4th ed.

\bibitem{Refregier2005}
P.~R\'{e}fr\'{e}gier, \enquote{Polarization degree of optical waves with
  non-gaussian probability density functions: Kullback relative entropy-based
  approach,} {\protect\JournalTitle{Opt. Lett.}} \textbf{30}, 1090--1092
  (2005).

\bibitem{kagalwala2013bell}
K.~H. Kagalwala, G.~Di~Giuseppe, A.~F. Abouraddy, and B.~E. Saleh,
  \enquote{Bell's measure in classical optical coherence,}
  {\protect\JournalTitle{Nature Photonics}} \textbf{7}, 72 (2013).

\bibitem{Yang2015}
S.~Yang, S.~A. Ponomarenko, and Z.~D. Chen, \enquote{Coherent pseudo-mode
  decomposition of a new partially coherent source class,}
  {\protect\JournalTitle{Opt. Lett.}} \textbf{40}, 3081--3084 (2015).

\end{thebibliography}
\bibliographyfullrefs{References}
\end{document}